\documentclass[runningheads]{llncs}

\usepackage{xspace}

\newcommand{\ie}{i.e.\xspace}
\newcommand{\eg}{e.g.\xspace}

\usepackage[T1]{fontenc}
\usepackage{graphicx}
\begin{document}
\title {Team Westwood Solution for MIDOG 2025 Challenge: An Ensemble-CNN-Based Approach for Mitosis Detection and Classification}

\titlerunning{MIDOG 2025 Team Westwood}
%
\author{
Tengyou Xu\inst{1} \and
Haochen Yang\inst{1} \and
Xiang `Anthony' Chen\inst{1} \and
Hongyan Gu \inst{2}\thanks{Correspondence: hgu2@kumc.edu} \and
Mohammad Haeri\inst{2}}
\authorrunning{Xu et al.}
%
\institute{
Department of Electrical and Computer Engineering, University of California Los Angeles, USA \and
Department of Pathology and Laboratory Medicine, University of Kansas Medical Center, USA}
\maketitle              
\

\begin{abstract}
This abstract presents our solution (Team Westwood) for mitosis detection and atypical mitosis classification in the \textbf{MItosis DOmain Generalization (MIDOG) 2025 challenge} \cite{ammeling_2025_15077361}. For mitosis detection, we trained an nnUNetV2 for initial mitosis candidate screening with high sensitivity, followed by a random forest classifier ensembling predictions of three convolutional neural networks (CNNs): EfficientNet-b3, EfficientNet-b5, and EfficientNetV2-s. For the atypical mitosis classification, we trained another random forest classifier ensembling the predictions of three CNNs: EfficientNet-b3, EfficientNet-b5, and InceptionV3. On the preliminary test set, our solution achieved an F1 score of \textbf{0.7450} for track 1 mitosis detection, and a balanced accuracy of \textbf{0.8722} for track 2 atypical mitosis classification. On the final test set, our solution achieved an F1 score of \textbf{0.6972} for track 1 mitosis detection, and a balanced accuracy of \textbf{0.8242} for track 2 atypical mitosis classification.

\end{abstract}
\section{Introduction}
In pathology, mitosis activity assessment in the Hematoxylin and Eosin (H\&E) slides by human pathologists can be challenging due to its small size and low prevalence in low-grade tumors \cite{meyer_breast_2005,bertram_computerized_2020}. Recent advancements in digital pathology and artificial intelligence (AI)  can provide a low-cost computer-assisted solution for more timely and precise examination \cite{bertram2022computer,gu2023augmenting}. Despite this, perhaps one hurdle for AI applicability is its generalizability on high variance of pathology datasets, due to three factors:  (1) the intrinsic appearance difference of mitosis and their mimickers across tumor types; (2) processing protocols from different labs; and (3) scanner imaging settings and image post-processing algorithms.

To fill this gap, several large-scale mitosis datasets covering various organs, scanners, and atypical mitotic figures have been recently curated and made publicly available \cite{bertram2019large,aubreville2020completely,aubreville2023comprehensive,shen2024deep}. Therefore, re-training AI models on these new datasets and running a more comprehensive evaluation has become increasingly necessary. In MIDOG 2022, we employed an EfficientNet-b3 CNN for both detection and classification of mitosis \cite{gu2022detecting}. While this design was compact in terms of model parameters, it relied on calculating attentions for mitosis localization, which could not be easily parallelized and thus had limited efficiency.

As an improvement, in MIDOG 2025 \cite{ammeling_2025_15077361}, we adopted the latest nnUNetV2\footnote{\url{https://github.com/MIC-DKFZ/nnUNet/tree/master/nnunetv2}}\cite{isensee2021nnu} as a lightweight and fast mitosis candidate localization and screening module. During training, both true positives and hard negatives were treated as positive samples to enhance its sensitivity. For each mitosis candidate, we then applied a ``heavier'' random forest of three CNN models (\ie EfficientNet-b3, EfficientNet-b5, and EfficientNetV2-s) to achieve specificity. For track 2 atypical mitosis classification task, we also used a random forest ensembling EfficientNet-b3, EfficientNet-b5, and InceptionV3, aiming to achieve more robust performance.

\section{Methods}
\subsection{Track 1: Mitosis detection} 
\subsubsection{AI Pipeline} 
\begin{figure*}
    \centering
    \includegraphics[width=1.0\linewidth]{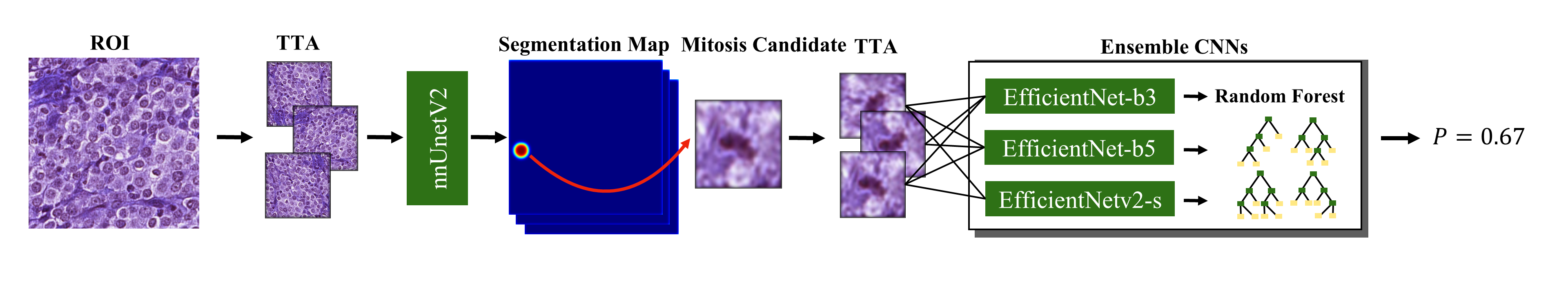}
    \caption{Illustration of our mitosis detection pipeline for track 1 challenge. ROI: region of interest, TTA: test-time augmentation, CNN: convolution neural network.}
    \label{fig:fig1}
\end{figure*}

Following the popular solutions in MIDOG 2022 \cite{aubreville2024domain}, we designed a two-stage mitosis segmentation-- verification pipeline to balance inferencing efficiency and detection performance. Specifically, we used the nnUNetV2 for stage-1 segmentation and a random forest of three CNNs for stage-2 verification, as shown in Figure~\ref{fig:fig1}.

\subsubsection{Dataset} We included MIDOG++\cite{aubreville2023comprehensive}, MITOS\_WSI\_CMC\cite{aubreville2020completely}, and MITOS\_C CMCT\cite{bertram2019large} for model training and validation (70,724 mitoses in total). Approximately $\sim$90\% of the slides or regions of interest (ROIs) were used for model training, and the rest for validation. To train the nnUNetV2, we cropped 253,703 (512$\times$512-pixel) patches from the training slides/ROIs, including positive patches randomly cropped around ground truth labels and negative patches randomly cropped from background. Both ground-truth mitoses and hard-negative mimickers were treated as positives (to improve the sensitivity, for nnUNetV2 training only). For each positive, we synthesized the segmentation mask by drawing a filled circle (45-pixel radius) centered at its location. The trained nnUNetV2 with the best sensitivity was then applied to both training and validation slides/ROIs. From all segmentation hotspot centroids, we extracted 140$\times$140-pixel patches (141,224 positives and 2,044,045 negatives\footnote{{The positives consist of 70,971 samples from ground-truth, 67,206 true positive predictions, and 3,047 false negatives by nnUNetV2. The negatives are false positive predictions by nnUNetV2.}}) for subsequent CNN training.

\begin{figure}
    \centering
    \includegraphics[width=1.0\linewidth]{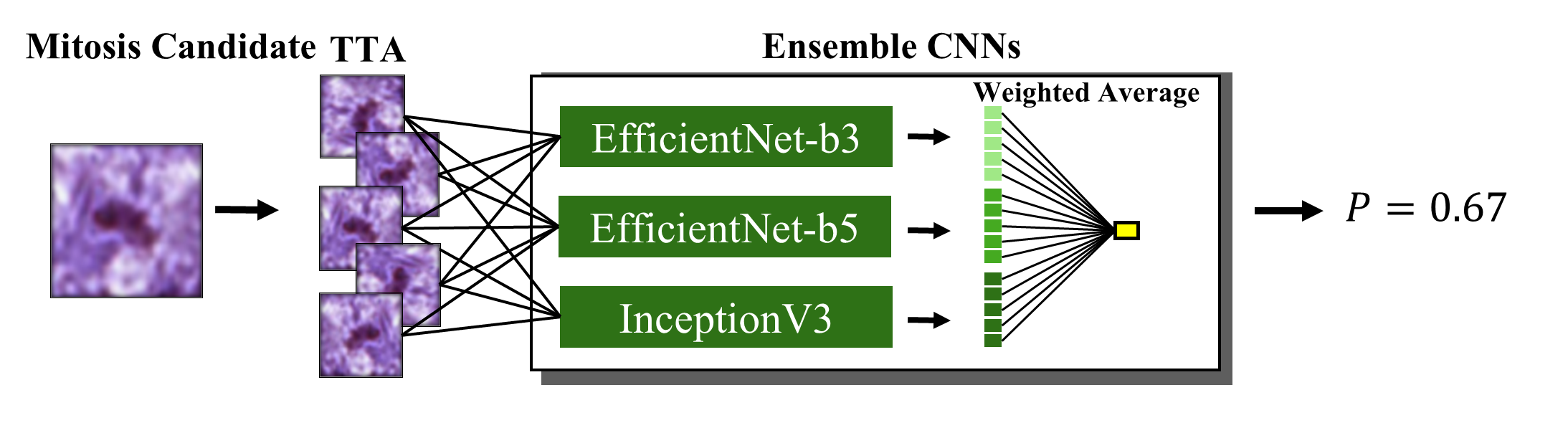}
    \caption{Illustration of track 2 atypical mitosis classification challenge.}
    \label{fig:fig2}
\end{figure}

\subsubsection{Model Training and Validation} Firstly, the segmentation model (nnUNetV2) was trained with oversample foreground percent 50\%, initial learning rate 0.001, weight decay $10^{-4}$, AdamW optimizer, DICE loss, Cosine Annealing LR with Warm Restarts (\texttt{T\_0}: 10, \texttt{T\_{mult}}: 1) for 50 epochs. Data augmentation included random image transform (\eg  crop, scaling, rotation, flip, mirror), color intensity transform (\eg brightness, contrast, and gamma adjustments), random gaussian noise and gaussian blur. After each training epoch, the checkpoint was evaluated on the entire validation set of slides/ROIs, and sensitivity was calculated. The checkpoint with the highest sensitivity was selected for final inferencing. 

For the classification models (multiple CNNs) training, we used the initial learning rate $8\times10^{-4}$, weight decay $10^{-4}$, AdamW optimizer, cross entropy loss, Cosine Annealing LR with Warm Restarts (\texttt{T\_0}: 15, \texttt{T\_{mult}}: 1) for 80 epochs. Data augmentation includes random image transform (\eg crop, flip, rotation), color adjustment (\eg brightness, hue, saturation), random gaussian noise and gaussian blur. We tried eight CNN variants: EfficientNet-b3, EfficientNet-b5, EfficientNetV2-s, EfficientNetV2-m, InceptionV3, ResNeXt50\_32x4d, ViT-b, and SwinV2-s. After each epoch, the checkpoint of each CNN was evaluated on the extracted validation patches. We trained each CNN for 80 epochs, and the top three CNNs (\ie EfficientNet-b3, EfficientNet-b5, and EfficientNetV2-s) with the highest F1 scores were selected to construct the final ensemble.

\subsubsection{Inferencing and Ensembling Training} Test-time augmentation (TTA) was applied to both nnUNetV2 ($\times$3; random flip and rotation) and each of the three CNNs ($\times$3; central random crop, flip, and rotation). For each candidate prediction, this TTA generated nine probability outputs (3 CNNs $\times$ 3 TTA). A random forest classifier (\texttt{n\_estimators}=260, \texttt{max\_depth}=1) was then trained using the output probabilities from all TTA results for each candidate as input features to estimate the final prediction probability. For submission, the pipeline was run on the test images using a 512-pixel sliding-window with 256-pixel overlap.

\subsection{Track2: Atypical mitosis classification}
\subsubsection{AI Pipeline} A random forest ensembe of three CNNs (see Figure \ref{fig:fig2}) was used to improve performance due to the small training set.

\subsubsection{Dataset} AMi-Br \cite{bertram2025histologic} and MIDOG 2025 Atypical Training Set \cite{weiss_2025_15188326} (13,077 mitoses and 2,580 atypical mitoses) were included. Approximately 85\% of the dataset was used for model training, and the rest for validation and threshold selection. All images were rescaled to 128$\times$128 pixels for training and inferencing.

\subsubsection{Model Training and Validation}
Similar to track 1-CNN, we trained eight CNN variants with the same hyperparameters and data augmentation strategy: EfficientNet-b3, EfficientNet-b5, EfficientNetV2-s, EfficientNetV2-m, Inception\_V3, ResNeXt50\_32x4d, ViT-b, and SwinV2-m. Three CNNs of EfficientNet-b3, EfficientNet-b5, and Inception\_V3 were selected because they achieved the highest balanced accuracies during validation.

\subsubsection{Inferencing and Ensembling Training} For each CNN, TTA ($\times$5; random flip and rotation) was used during the inferencing. An ensemble module made the final prediction by averaging the 15 concatenated probability outputs with equal weights, as this approach outperformed the random forest method in the Track 2 preliminary test.

\section{Results}
\paragraph{Preliminary test phase}
In track 1, our approach achieved an overall mitosis detection F1 score of 0.7450 (ranked at \#19), which is 2.9\% lower than the baseline method (F1: 0.7672). The per-tumor F1 scores were 0.8462 (tumor 1), 0.6861 (tumor 2), 0.7601 (tumor 3), and 0.8000 (tumor 4), respectively. Upon further inspection, our approach achieved a relatively low recall in tumor 2 (0.5839), which in turn resulted in lower overall performance.

For track 2, our approach achieved balanced accuracy of 0.8722 for atypical mitosis classification, which is 9.9\% higher than the baseline approach (0.7933).

\paragraph{Final test phase}
In track 1, our approach achieved an overall mitosis detection F1 score of 0.6972 (ranked at \#7), which is 1.3\% higher than the baseline method (F1: 0.6883). The F1 scores for Hotspot ROIs, Random ROIs, and Challenging ROIs were respectively 0.7318, 0.6553, and 0.5281, while the per-tumor F1 scores 0.7229 (tumor 1), 0.4154 (tumor 2), 0.7992 (tumor 3), 0.6740 (tumor 4), 0.6897 (tumor 5), 0.6947 (tumor 6), 0.6876 (tumor 7), 0.7703 (tumor 8), 0.6512 (tumor 9), 0.7300 (tumor 10), 0.4186 (tumor 11), and 0.3873 (tumor 12)

For track 2, our approach achieved balanced accuracy of 0.8242 for atypical mitosis classification, which is 0.4\% lower than the baseline approach (0.8274).

\section{Discussion}
The results demonstrated that an ensembling module using random forest can improve the robustness and generalizability of AI detection. During our validation stage, we observed improvements in F1 score robustness when the number of trees (\textit{i.e.,} \texttt{n\_estimators}) increased from 100 to 500. The result suggests that incorporating the ensembling approach can enable the mitosis detection pipeline to be less subject to the threshold selection across different organs, patients, and scanners during practical usage.

Based on the result, we suggest two directions for future improvements:

\begin{enumerate}
    \item \textbf{Data quality}. It is noteworthy that all current public mitosis detection dataset was scanned using a single $z$ layer. In the meantime, Z-stacked scanning, which has become more available recently, can improve the whole slide imaging quality by preserving additional depth information in the $z$ direction. In our previous study \cite{gu2025zstackscanningimproveai}, we observed that using z-stacked scans of five layers improved AI mitosis detection recall by 17.14\%, while only having a marginal impact on the precision. Future work may explore similar techniques to improve the whole slide imaging quality (in both $x-y$ and $z$ directions) and measure the corresponding effectiveness.
    
    \item \textbf{Data quantity and variation.}
    During the final test phase of track 1, we noticed that the F1 score achieved by the 3$^{rd}$ solution (0.7085) to 10$^{th}$ (0.6883) only varied by $\sim 2\%$. Therefore, modifications on AI models or detection pipelines may not be the most effective way to further improve the performance: the focus may shift to extend the training dataset, such as incorporating additional hard negative samples, to further improve the model’s capability to distinguish the challenging, confusing patterns.
\end{enumerate}

%
%
%
\bibliographystyle{splncs04}
%
\bibliography{ref}

\end{document}